\documentclass[a4paper]{jpconf}

\usepackage{amssymb}
\usepackage{epsfig}

\def\C{\mathcal{C}}

\def\Epsilon{\mathrm{E}}
\def\Kappa{\mathrm{K}}

\def\PBG#1#2{\{#1, #2\}^G}
\def\PBGamma#1#2{\{#1, #2\}^\Gamma}

\def\PBA#1#2{\{#1, #2\}^A}
\def\PBKappa#1#2{\{#1, #2\}^{\Kappa}}
\def\PBAlpha#1#2{\{#1, #2\}^\Alpha}

\def\H{\mathcal{H}}

\def\Alpha{\mathrm{A}}
\def\kuchar{Kucha\v{r}}

\begin{document}
\title{Towards conformal loop quantum gravity}

\author{Charles H-T Wang}

\address{School of Engineering and Physical Sciences,
University of Aberdeen, King's College, Aberdeen AB24 3UE,
UK\\
Space Science and Technology Department, Rutherford Appleton
Laboratory, Didcot, Oxon OX11 0QX, UK}

\ead{c.wang@abdn.ac.uk}

\begin{abstract}
A discussion is given of recent developments in canonical gravity
that assimilates the conformal analysis of gravitational degrees
of freedom. The work is motivated by the problem of time in
quantum gravity and is carried out at the metric and the triad
levels. At the metric level, it is shown that by extending the
Arnowitt-Deser-Misner (ADM) phase space of general relativity
(GR), a conformal form of geometrodynamics can be constructed. In
addition to the Hamiltonian and diffeomorphism constraints, an
extra first class constraint is introduced to generate conformal
transformations. This phase space consists of York's mean
extrinsic curvature time, conformal three-metric and their
momenta. At the triad level, the phase space of GR is further
enlarged by incorporating spin-gauge as well as conformal
symmetries. This leads to a canonical formulation of GR using a
new set of real spin connection variables. The resulting
gravitational constraints  are first class, consisting of the
Hamiltonian constraint and the canonical generators for spin-gauge
and conformorphism transformations. The formulation has a
remarkable feature of being parameter-free. Indeed, it is shown
that a conformal parameter of the Barbero-Immirzi type can be
absorbed by the conformal symmetry of the extended phase space.
This gives rise to an alternative approach to loop quantum gravity
that addresses both the conceptual problem of time and the
technical problem of functional calculus in quantum gravity.
\end{abstract}

\section{Introduction}

The problem of time in attempts to quantize general relativity
(GR) has been examined extensively over the past decades. An
excellent review is provided by Isham in \cite{Isham1993}.
\kuchar{} has proposed to deparametrize GR by finding a set of
four spacetime scalars constructed from the gravitational
variables as privileged spacetime coordinates with respect to
which other gravitational variables evolve \cite{Kuchar1971}. This
prescription has found applications in certain $(1+1)$-dimensional
midisuperspace models \cite{Torre1999} but its success is
otherwise limited \cite{TorreVaradarajan1999}. In classical GR,
the issue of time also arises as part of the initial value
problem. In this regard, a resolution has long been found by York
in terms of the conformally invariant decomposition of tensors on
the spatial hypersurface \cite{York1971, York1972}. The mean
extrinsic curvature is identified as time that traces the
evolution of conformal 3-geometry that represents the dynamical
degrees of freedom of gravity as nonlinear gravitational waves.
York's mean curvature time is defined with respect to a foliation
and therefore is not a spacetime scalar. Consequently, it cannot
be a time coordinate for deparametrized GR in the sense of
\kuchar{} \cite{Kuchar1992a}. Furthermore, the original dynamical
formulation of GR using the York time is not Hamiltonian
\cite{ChoquetBruhatYork1980}. It is therefore not clear how this
description can be employed in the canonical quantum gravity
programme. Although by restricting the spacetime foliation to that
having a constant mean curvature over the spatial hypersurface, a
canonical description could be formulated. It comes at the price
of an implicit Hamiltonian whose quantization is not well
understood \cite{FischerMoncrief1997, Carlip2001}. Alternatively
one might consider a nonlinear quantization scheme
\cite{Wang2005a}, but even so the potential loss of general
covariance is of concern.

These issues have been addressed in the recent papers
\cite{Wang2005b, Wang2005c}, in which a class of new canonical
structures of GR have been found to accommodate the York time. The
essential methodology involve extending the Arnowitt-Deser-Misner
(ADM) phase space of GR to allow for a new first class constraint
that generates conformal transformations \cite{Wang2005b}. This
way, it is possible to formulate the canonical evolution of the
3-geometry along with to York's mean curvature time
\cite{Wang2005b}.  A similar procedure can be applied to the triad
formalism of canonical gravity. This leads to the canonical
evolution of conformal triad along with the York time. The
resulting canonical formulation can be further turned into a
spin-gauge type formulation of canonical gravity in a fashion of
Ashtekar \cite{Ashtekar1986} and Barbero \cite{Barbero1995a}.
Unlike the Ashtekar-Barbero real spin-gauge theory of GR, however,
the new canonical structure turns out to be free from a free
conformal parameter, called the Barbero-Immirzi parameter
\cite{Immirzi1997}. This is due to the presence of the conformal
symmetry in the extended phase space which is capable of absorbing
the conformal arbitrariness in defining the spin-gauge variables
\cite{Wang2005c}. This opens up a new approach to loop quantum
gravity that is free from the Barbero-Immirzi parameter while
having an explicit time variable in terms of the mean
curvature\cite{Wang2005c}. Below, we review the essential results
from \cite{Wang2005b, Wang2005c} with the aide of a diagram to
illustrate the hierarchy of canonical transformations and a table
to summarize the canonical variables being introduced. The
convention and definitions follow from \cite{Wang2005b,
Wang2005c}.

\section{Conformal geometrodynamics: the $\Gamma$-variables}

The standard geometrodynamics is based on the ADM variables of GR
consisting of the spatial metric $g_{a b}$ with conjugate momentum
$p^{ab}$. The ADM diffeomorphism (momentum) and Hamiltonian
constraints are given respectively by
\begin{eqnarray*}
\H_a = -2{p}^b{}_{a;b} \approx 0,
\quad
\H_\perp = G_{a b c d}\,
p^{a b} p^{c d} -  \mu R \approx 0
\end{eqnarray*}
where $\mu = \sqrt{\det g_{ab}}$ is the scale factor,
$G_{a b c d}$ is the DeWitt metric, $R$ is the Ricci scalar and the symbol
`$\approx$' denotes a weak equality {\it a la} Dirac. We will
perform a series of canonical transformations of the gravitational
variables will refer to each set of the canonical variables using
a capital letter. Thus, we refer to the ADM variables
$(g_{a b},p^{a b})$ as the `$G$-variables', with respect to which the
Poisson bracket is denoted by $\PBG{\cdot\,}{\cdot}$. We shall
also denote the ADM constraints by
\begin{eqnarray*}
\C^G_a := \H_a \approx 0, \quad \C^G_\perp := \H_\perp  \approx 0 .
\end{eqnarray*}
In \cite{Wang2005b} the ADM phase space has been extended to that
consisting of York's time $\tau := (4/3) K$, where $K$ is the mean extrinsic curvature,
with
$\mu$ as momentum and conformal metric $\gamma_{a b}$ with
momentum $\pi^{ab}$. Based on York's decomposition of tensors, a
canonical transformation has been found to relate the
$G$-variables to the `$\Gamma$-variables' $(\gamma_{a b}, \pi^{a
b}; \tau, \mu)$ via
\begin{eqnarray*}
g_{a b} &=& \phi^4 \gamma_{a b}, \;\;
p^{a b} = \phi^{-4}\pi^{a b} - \frac12  \, \phi^2 \bar{\mu}\,
\gamma^{ab}\tau
\end{eqnarray*}
where $\bar{\mu} := \sqrt{\det \gamma_{ab}}$ and
$\phi := ({\mu}/{\bar{\mu}})^{1/6}$ are called the conformal scale
factor and conformal
factor respectively. It can be seen from the above relations that
a local rescaling of
$\gamma_{ab}$ and $\pi^{a b}$
while holding $\tau$ and $\mu$ leaves ${g}_{a b}, p^{a b}$ invariant.
This redundancy of the $\Gamma$-variables is
compensated by the `conformal constraint':
\begin{eqnarray*}
\C^\Gamma := {\gamma}_{ab}\pi^{ab}
\approx 0
\end{eqnarray*}
that also
generates conformal transformations through Poisson bracket
with respect to the $\Gamma$-variables denoted by
$\PBGamma{\cdot\,}{\cdot}$.

From the ADM constraints, the diffeomorphism and Hamiltonian
constraints for the $\Gamma$-variables are obtained as
\begin{eqnarray*}
\label{C^Gamma_a}
\C_a^\Gamma &:=& \tau_{,a} {\mu} -2 \pi^b{}_{a;b} \approx 0
\\
\C_\perp^\Gamma &:=& -\frac{3}{8}\, \tau^2 \mu +
\frac1{\mu}\,\pi{}_{ab}\pi^{ab}
-\mu R
\approx 0
\label{H^Gamma}
\end{eqnarray*}
respectively. Using the preservation of the
Poisson bracket relations by
the canonical transformation from the $G$- to $\Gamma$-variables,
the constraints
$C^\Gamma, \C_a^\Gamma$ and $\C_\perp^\Gamma$
can be shown to be of first class \cite{Wang2005b, Wang2005c}.

\section{Triad formalism: the $K$-variables}

The triad formalism of canonical GR is obtained by
introducing the triad $e^i_a$ with inverse $e^a_i$
and densitized triad $E_i^a = \mu\, e_i^a$
with inverse $E^i_a$
so that the 3-metric takes the form
\begin{eqnarray*}
g_{a b} = \mu^2  E^i_a  E^i_b
\end{eqnarray*}
using
the triad (spin) indices $i, j, \cdots = 1, 2, 3$.

The
orientation is chosen so that $\det e^i_a > 0$. The
spin-valued extrinsic curvature $K^i_{a}$ is defined such that the extrinsic
curvature tensor becomes
\begin{eqnarray*}
{K}_{ab} = \frac\mu2 \, {K}^i_{(a} E^i_{b)} .
\end{eqnarray*}
The quantities $(K^i_{a}, E_i^a)$ form
coordinates for an extended phase space of
GR and are referred to as
the `$K$-variables' in our discussion.
The redundancy due to spin transformations, i.e. local changes of the triad frame,
in this description is eliminated by means of the spin constraint:
\begin{eqnarray*}
\C_i^K := \epsilon^{}_{i j k} K^{}_{a j} E^a_{k}
\approx
0 .
\end{eqnarray*}


\section{Spin-gauge formalism: the $A$-variables}

The canonical structure of GR based on which the present
loop quantum gravity is formulated using
the one-parameter family of phase spaces
given by:
\begin{eqnarray*}
A^i_a &:=& \Gamma^i_a + \beta \, K^i_a,\;\; P_i^a :=
\frac{E_i^a}{\beta}
\end{eqnarray*}
parametrized by a nonzero complex constant $\beta$. Here
$\Gamma^i_a$ is the Levi-Civita spin connection
while $A^j_a$ is a spin connection with torsion.
The spin covariant derivative and curvature 2-form
associated with $A^j_a$
are
denoted by $D$ and $F^{k}_{a b}$ respectively.
We refer to
the pairs $(A^i_a, P_i^a)$ as the `$A$-variables' and denote the
corresponding PB by $\PBA{\cdot\,}{\cdot}$.
The transformation
from the $K$- to $A$-variables is canonical and is Poisson bracket preserving.
In the $A$-variables the spin constraint can be expressed in the `Gauss form' as follows:
\begin{eqnarray*}
\C^A_i := D_a P^a_i = \C^K_i \approx 0 .
\end{eqnarray*}
Further, the diffeomorphism constraint takes the form
\begin{eqnarray*}
\C^A_a &:=& F^k_{ab} P_k^b - A^k_{a} \C^A_k  \approx 0 .
\end{eqnarray*}
The constraints $\C^A_k$ and $\C^A_a$ respectively generates
rotations and diffeomorphisms through their Poisson brackets with the
$A$-variables. The Hamiltonian constraint then becomes:
\begin{eqnarray*}
\C_\perp^A &:=& \frac{1}{\mu} \left[ \epsilon_{i j k}\, \beta^2
F^k_{a b} -\frac{4\beta^2+1}{2}\,\tilde{K}^i_{[a} \tilde{K}^j_{b]}
\right] P^a_i P^b_j
\approx 0 .
\end{eqnarray*}
The constraints $\C^A_k, \C_a^A$ and  $\C_\perp^A$ hence form a
set of independent first class constraints. Ashtekar's original
gauge formalism of GR corresponds to the choice $\beta=\pm i/2$ so
as to obtain a polynomial expression \cite{Ashtekar1986}. In
Barbero's modified approach, $\beta$ is considered as a real and
positive parameter in order to resolve the reality problem in loop
quantum gravity \cite{Barbero1995a}.

\section{Conformal triad formalism: the $\Kappa$-variables}

The parameter $\beta$
is an arbitrary scaling factor in
defining the $A$-variables.
If an alternative set of spin-gauge
variables can be found that possesses a conformal symmetry,
then such an arbitrariness may be absorbed.
To proceed, we introduce the conformal triad $\bar{e}^i_a$ with
inverse $\bar{e}_i^a$ so that
$\gamma_{a b} = \bar{e}^i_a\bar{e}^i_b$ and
$\gamma^{a b} = \bar{e}_i^a \bar{e}_i^b$.
Further, we introduce the densitized triad
$\Epsilon_i^a = \bar{\mu}\, \bar{e}_i^a$ with inverse
$\Epsilon_i^a = \bar{\mu}^{-1}\, \bar{e}^i_a$. The Levi-Civita spin connection of
$\gamma_{a b}$ will be denoted by $\bar{\Gamma}$ and the
associated covariant differentiation also denoted by $\bar{\nabla}$ and
`$|$'.
A trace-split of the extrinsic curvature $K^i_a$ is then performed
in a conformally invariant manner. These considerations lead to:
\begin{eqnarray*}
E_i^a &=& \phi^{4}\, \Epsilon_i^a, \;\;
K^i_a = \phi^{-4}\, \Kappa^i_a + \frac12\,\phi^2 \bar{\mu}\,
\Epsilon^i_a \tau
\end{eqnarray*}
where we have introduced $\Kappa^i_a$ to function as the
`conformal extrinsic curvature'. We have thus arrived at a set
of conformal triad description of GR using $(\Kappa^C,
\Epsilon_C):=(\Kappa^i_a, \Epsilon^a_i; \tau, \mu)$, called the
`$\Kappa$-variables'. Using the corresponding Poisson bracket denoted by
$\PBKappa{\cdot\,}{\cdot}$, we can show that
the transformation
from the $K$- to $\Kappa$-variables is canonical and is Poisson bracket preserving.

\section{Conformal spin-gauge formalism: the $\Alpha$-variables}

The canonical
transformation from the $K$- to $A$-variables can be
derived by adding
the total divergence
$(1/\beta)E_k^c \dot{\Gamma}_{c}^{k}$ to the canonical action of GR,
for an arbitrary nonzero constant coefficient $\beta$. By analogy, we
can add the total divergence
$({1}/{\alpha})\Epsilon_k^c \dot{\bar{\Gamma}}_{c}^{k}$
to the time-derivative terms of the canonical action of GR in the
$\Kappa$-variables:
\begin{eqnarray*}
\mu\,\dot{\tau} + \Epsilon_i^a\dot{\Kappa}^i_a + \frac{1}{\alpha}
\Epsilon_k^c \dot{\bar{\Gamma}}_{c}^{k} = \mu\,\dot{\tau} +
\frac{\Epsilon_i^a}{\alpha}   (\bar{\Gamma}^i_a +
\alpha\Kappa^i_a)\,\dot{}
\end{eqnarray*}
for any constant ${1/\alpha}$.
However, for any positive $\alpha$ the conformal symmetry of the
$\Kappa$-variables can absorb this constant by redefining the
variables as follows
\begin{eqnarray*}
\Epsilon_i^a\rightarrow \alpha \Epsilon_i^a,\;
\Kappa^i_a\rightarrow \frac{\Kappa^i_a}{\alpha},\;
\bar{\Gamma}^i_a\rightarrow \bar{\Gamma}^i_a .
\end{eqnarray*}
In this case,
time-derivative terms above become
$\mu\,\dot{\tau}+\Pi_i^a\dot{\Alpha}^i_a$ in a new set of
variables \cite{Wang2005c}:
\begin{eqnarray*}
\Alpha^i_a &:=& \bar{\Gamma}^i_a + \Kappa^i_a, \;\; \Pi_i^a :=
\Epsilon_i^a .
\end{eqnarray*}
We call $(\Alpha^C, \Pi_C):=(\Alpha^i_a, \Pi_i^a; \tau, \mu)$ the
`$\Alpha$-variables' and denote the associated Poisson bracket by
$\PBAlpha{\cdot\,}{\cdot}$. The transformation
from the $\Kappa$- to $\Alpha$-variables is canonical and is Poisson bracket preserving.

The spin covariant derivative associated with $\Alpha^i_a$ and its
curvature 2-form are denoted by $\bar{D}$ and $\bar{F}$
respectively.  It
follows that the conformal, spin and diffeomorphism constraints
in the $\Alpha$-variables becomes
\begin{eqnarray*}
\C^{\Alpha} &:=& \frac{1}{2}\, \Kappa^i_{a} \Pi^a_i
\approx 0
\\
\C^{\Alpha}_i
&:=&
\bar{D}_a \Pi^a_i
\approx 0
\\[1.5mm]
\C^\Alpha_a &:=& \tau_{,a} {\mu} + \bar{F}^k_{ab} \Pi_k^b -
\Alpha^k_{a} \C^\Alpha_k
\approx 0
\end{eqnarray*}
they generate the
corresponding transformations using the Poisson bracket $\PBAlpha{\cdot\,}{\cdot}$.
Finally, the Hamiltonian constraint becomes
\begin{eqnarray*}
\C_\perp^\Alpha &:=& - \frac{3}{8}\, \tau^2 \mu + 8\mu\, \phi^{-5}
\bar{\Delta} \phi \nonumber
+ \frac{1}{\mu} \left[ \phi^8 \epsilon_{kij} \bar{F}^k_{ab} -
\frac{4\phi^8+1}{2} \Kappa^i_{[a}  \Kappa^j_{b]} \right] \Pi_i^a
\Pi_j^b
\approx 0
\end{eqnarray*}
where $\bar{\Delta} := \gamma^{ab}\bar{\nabla}_a\bar{\nabla}_b$ is
the Laplacian associated with the conformal metric $\gamma_{ab}$.
In the last equation above, the third term is similar to the
standard case with $\beta\rightarrow \phi^4$. There, the
`additional' first term is due to the York time $\tau$ being
separated from the conformal part of kinematics whereas the second
term counts for the conformal factor $\phi$ being a local function
of the $\Alpha$-variables.

By using the preservation of the Poisson bracket of $\H_\perp(x)$ and
$\H_\perp(x')$ throughout the canonical transformations discussed above,
one sees that $\C_\perp^\Alpha$ and the canonical
generators $\C^{\Alpha}$, $\C^{\Alpha}_i$ and $\C^\Alpha_a$ form a
set of first class constraints for the above conformal spin-gauge
formulation of GR using the $\Alpha$-variables \cite{Wang2005c}.

\section{Conclusions}

We list all canonical variables consider here in table~\ref{tab_vars}.
\begin{table}[ht!]
\caption{\label{tab_vars}Canonical variables}
\begin{center}
\begin{tabular}{l@{\qquad}c@{\qquad}c@{\qquad}c@{\qquad}c@{\qquad}c@{\qquad}c}
\br
Variable set& $G$ & $K$ & $A$ & $\Gamma$ & $\Kappa$ & $\Alpha$ \\
\mr
Field variables&$g_{a b}$&$K^i_a$&$A^i_a$&$\gamma_{a b}; \tau $&$\Kappa^i_a; \tau$&$\Alpha^i_a; \tau$\\[1mm]
Momentum variables&$p^{a b}$&$E_i^a$&$P_i^a$&$\pi^{a b}; \mu$     &$\Epsilon_i^a; \mu$&$\Pi_i^a; \mu$\\
\br
\end{tabular}
\end{center}
\end{table}

\begin{figure}[ht!]
\begin{center}
\epsfig{file=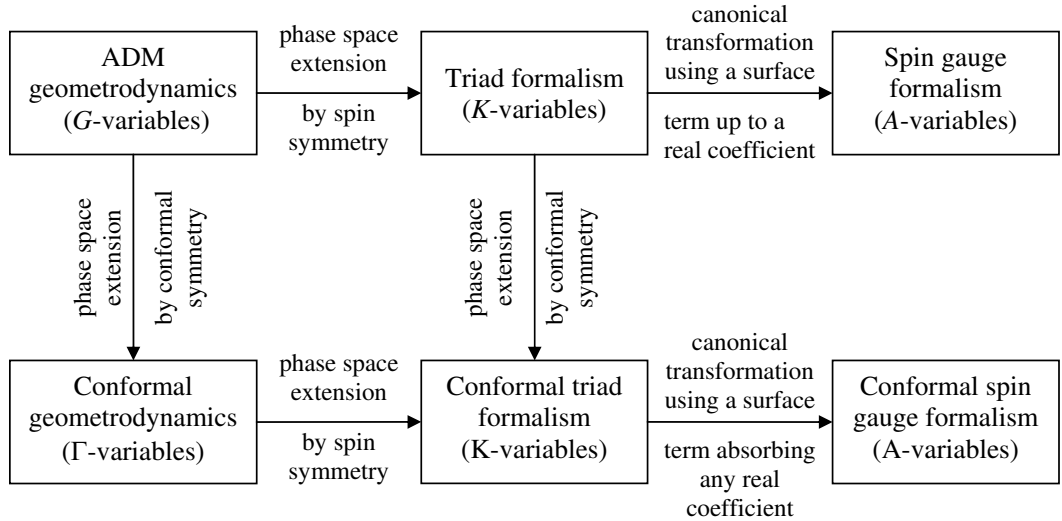,width=155mm,angle=0}
\end{center}
\caption{\label{fig_cano}Hierarchy of phase variable sets in
canonical GR}
\end{figure}
Through a sequence of canonical transformations as summarized in figure~\ref{fig_cano}, a new canonical
structure of GR has been found with the following features:

\begin{enumerate}

\item No preferred spacetime foliation is required.

\item Lie symmetries incorporating:
\begin{enumerate}
    \item Diffeomorphism -- as the essential translational invariance.
    \item Spin-gauge -- needed for the Yang-Mills gauge treatment and the use of the loop representation.
    \item Conformal invariance -- for a theory free from the Barbero-Immirzi parameter that also
    addresses the problem of time and  unitary quantum evolution.
\end{enumerate}

\item The entire system of constraints is first class.

\end{enumerate}

As future work, it may be possible to obtain a conformal
version of the spin network used in the present loop quantum gravity. The
obtained `conformal loop quantum gravity' may be applied to,
e.g. the black hole entropy calculation. In the absence of the
Barbero-Immirzi parameter such a result should be of interest.

\ack{I am most grateful to
J F Barbero,
R Bingham,
S Carlip,
A E Fischer,
L Griguolo,
G Immirzi,
C J Isham,
G Mena Marugan,
J T Mendon\c{c}a,
N O'Murchadha
and
J W York for inspiring discussions and to
the Aberdeen Centre for Applied Dynamics Research and the
CCLRC Centre for Fundamental Physics for financial support.}

\end{document}